\documentclass[aps,prl,twocolumn,superscriptaddress,showpacs,preprintnumbers,amsmath,amssymb]{revtex4-2}
\usepackage{graphicx} 
\usepackage{dcolumn}  

\graphicspath{{ps}}

\usepackage{orcidlink}
\usepackage{multirow,tabularx}

\begin{document}

\title{ \quad\\[1.0cm] Evidence of $h_{b}(\text{2P}) \to \Upsilon(\text{1S})\eta$ decay and search for $h_{b}(\text{1P,2P}) \to \Upsilon(\text{1S})\pi^0$ with the Belle detector}

\noaffiliation
  \author{E.~Kovalenko\,\orcidlink{0000-0001-8084-1931}} 
  \author{I.~Adachi\,\orcidlink{0000-0003-2287-0173}} 
  \author{H.~Aihara\,\orcidlink{0000-0002-1907-5964}} 
  \author{D.~M.~Asner\,\orcidlink{0000-0002-1586-5790}} 
  \author{T.~Aushev\,\orcidlink{0000-0002-6347-7055}} 
  \author{R.~Ayad\,\orcidlink{0000-0003-3466-9290}} 
  \author{V.~Babu\,\orcidlink{0000-0003-0419-6912}} 
  \author{Sw.~Banerjee\,\orcidlink{0000-0001-8852-2409}} 
  \author{K.~Belous\,\orcidlink{0000-0003-0014-2589}} 
  \author{J.~Bennett\,\orcidlink{0000-0002-5440-2668}} 
% ...
\collaboration{The Belle Collaboration}     

\begin{abstract}
We report the first evidence for the $h_{b}(\text{2P}) \to \Upsilon(\text{1S})\eta$ transition with a significance of $3.5$ standard deviations.
The decay branching fraction is measured to be
  $\mathcal{B}[h_{b}(\text{2P}) \to \Upsilon(\text{1S})\eta]=(7.1 ~^{+3.7} _{-3.2}\pm 0.8)\times10^{-3}$,
which is noticeably smaller than expected.
We also set upper limits on $\pi^0$ transitions of 
  $\mathcal{B}[h_{b}(\text{2P}) \to \Upsilon(\text{1S})\pi^0] < 1.8\times10^{-3}$,
    and
  $\mathcal{B}[h_{b}(\text{1P})\to \Upsilon(\text{1S})\pi^0] < 1.8\times10^{-3}$, at the $90\%$ confidence level.
  These results are obtained with a $131.4$~fb$^{-1}$ data sample collected near the $\Upsilon(\text{5S})$ resonance with the Belle detector at the KEKB asymmetric-energy $e^+e^-$ collider.

\end{abstract}


\maketitle


{\renewcommand{\thefootnote}{\fnsymbol{footnote}}}
\setcounter{footnote}{0}


The study of rare hadronic transitions between bottomonium states, constituted by a $b\bar{b}$ quark pair, provides important information about their structure and allows for precise tests of the effective field theories used to model non-perturbative QCD, such as the QCD multipole expansion~\cite{voloshin1979}.
For example, the $h_{b}(\text{2P}) \to \Upsilon(\text{1S})\eta$ decay is of great interest as its rate is suppressed by the heavy quark spin symmetry.
We expect the decay properties of spin-singlet $^1P_1$ states, $h_b(\text{1P})$ and $h_b(\text{2P})$~\cite{Belle:2011wqq}, to be similar to those of their spin-triplet partners, $\chi_{b1}(\text{1P})$ and $\chi_{b1}(\text{2P})$~\cite{Godfrey:2002rp}.
The authors in Ref.~\cite{Li:2012is} argue that the ratio $R_{h_b}$ of the annihilation rates for $h_b(\text{2P})$ and $h_b(\text{1P})$ is the same as the corresponding ratio $R_{\chi_{b1}}$ for $\chi_{b1}(\text{2P})$ and $\chi_{b1}(\text{1P})$.
However, the estimated value of $R_{h_b} / R_{\chi_{b1}}$ from measured decay rates to bottomonium (non-annihilation decays) is $0.25\pm0.25$~\cite{Li:2012is}.
Although this value may differ by $3.0$ standard deviations ($\sigma$) from unity, we estimate the difference to be $1.5\sigma$ if we use current branching fractions and correlations among the uncertainties, resulting in a value of $0.24^{+0.47}_{-0.24}$.
Nevertheless, the discrepancy would further increase if the rate of $h_{b}(\text{2P}) \to \Upsilon(\text{1S})\eta$ were as large as $10\%$, as the same authors predict based on the $\Upsilon(\text{3S})\to h_b(\text{1P})\pi^0$ decay rate.
Rates of the isospin-violating transitions $h_{b}(\text{1P},\text{2P}) \to \Upsilon(\text{1S})\pi^0$ are expected to be further suppressed~\cite{Voloshin:2007dx} and can be tested in the same final state as $h_{b}(\text{2P}) \to \Upsilon(\text{1S})\eta$.

In this Letter, we report studies of $h_b(\text{1P},\text{2P})$ hadronic transitions to the $\Upsilon(\text{1S})$ state with emission of either $\eta$ or $\pi^0$ using a sample of $121.4$~fb$^{-1}$ data and $12.0$~fb$^{-1}$ of energy-scan data.
These data were collected at and near the $\Upsilon(\text{5S})$ resonance, respectively, with the Belle detector~\cite{Abashian:2000cg, Belle:2012iwr} at the KEKB asymmetric-energy $e^+e^-$ collider \cite{Kurokawa:2001nw, Abe:2013kxa}.  
We perform a full reconstruction of the $e^+e^-\to h_b(\text{1P},\text{2P})\pi^+\pi^-$ process, with $h_b(\text{2P})\to\Upsilon(\text{1S})\eta(\pi^0)$ or $h_b(\text{1P})\to\Upsilon(\text{1S})\pi^0$, $\eta(\pi^0) \to \gamma\gamma$, and $\Upsilon(\text{1S}) \to \ell^+\ell^-$, where $\ell$ stands for an electron or a muon.
The signal yield is extracted by fitting the $M_{\gamma\gamma}$ vs. $M^{\text{rec}}_{\pi\pi}$ distribution, since for signal events, the diphoton invariant mass, $M_{\gamma\gamma}$, peaks at either the $\pi^0$ or the $\eta$ mass, and the mass of the system recoiling against two pions, $M^{\text{rec}}_{\pi\pi}$, peaks at the $h_b(\text{1P,2P})$ masses.
Here, the recoil mass is defined as $M^{\text{rec}}_{X}=\sqrt{s + M^2_{X} -2\sqrt{s}E^*_{X}}$, with $E^*_{X}$ being the energy of $X$ in the center-of-mass frame and $\sqrt{s}$ being the $e^+e^-$ collision invariant mass.


The Belle detector is a large-solid-angle magnetic spectrometer that consists of a silicon vertex detector, a $50$-layer central drift chamber (CDC), an array of aerogel threshold Cherenkov counters (ACC), a barrel-like arrangement of time-of-flight scintillation counters (TOF), and an electromagnetic calorimeter (ECL) comprised of CsI(Tl) crystals located inside a superconducting solenoid coil that provides a $1.5$ T magnetic field. 
An iron flux-return yoke located outside of the coil (KLM) is instrumented with resistive-plate chambers to detect $K_L^0$ mesons and muons.
The $z$-axis of the detector points in the direction opposite to the positron beam.
A detailed description of the Belle detector is given elsewhere~\cite{Abashian:2000cg}.

We optimize event selection using a Monte Carlo (MC) simulation.
The MC events are generated with EvtGen~\cite{Lange:2001uf}, and the detector response is modeled with GEANT3~\cite{Brun:1987ma}. 
The dynamics of the $e^+e^-\to h_b(\text{1P},\text{2P})\pi^+\pi^-$ process is modeled according to the measurement in Ref.~\cite{Bondar:2011aa}.
The dilepton decay of $\Upsilon(\text{1S})$ is simulated taking into account the proper spin dynamics.
Final-state radiation is simulated with the PHOTOS package~\cite{PHOTOS}.
The detector simulation incorporates trigger simulation, and the varying data-taking and beam conditions over time.

To identify dominant sources of background, we use two sets of MC events.
The first set corresponds to six times the integrated luminosity of the data and includes a variety of processes, such as $e^+e^-$ annihilation into $q\Bar{q}$ ($q=u$, $d$, $s$, $c$), $B$ meson pairs ($B^{(*)}_s\Bar{B}^{(*)}_s$, $B^{(*)}\Bar{B}^{(*)}(\pi)$), and known decays of the $\Upsilon(\text{5S})$.
The second set is equivalent to four times the integrated luminosity and  consists of background processes such as $e^+e^-\to e^+e^-$, $e^+e^-\to \mu^+\mu^-$, $e^+e^-\to e^+e^- q \Bar{q}$, and $e^+e^-\to \tau^+\tau^-$.

The event selection proceeds as follows: first, a $\pi^+\pi^-$ pair is selected using the same requirements as in Ref.~\cite{Belle:2012fkf}, since we use measured $h_b(\text{1P},\text{2P})$ signal yields for branching fraction calculation.     
We require $dr<0.3$~cm and $|dz|<2.0$~cm, where $dr$ and $dz$ are transverse and longitudinal impact parameters for the track with respect to the $e^+e^-$ interaction point.
Tracks are identified as pions using requirements $\frac{\mathcal{L}_{\pi}}{\mathcal{L}_K+\mathcal{L}_{\pi}} > 0.1$, $\frac{\mathcal{L}_{\pi}}{\mathcal{L}_p+\mathcal{L}_{\pi}} > 0.1$, and $\mathcal{P}_{e}<0.9$, where $\mathcal{L}_i$ is a particle identification likelihood for the track assigned based on CDC, ACC, TOF information, and $\mathcal{P}_{e}$ is a likelihood ratio based on CDC, ACC, and ECL information~\cite{Hanagaki:2001fz} that tests the electron hypothesis.
The pion identification efficiency is $99\%$, while electron and kaon misidentification rates are $8\%$ and $4\%$, respectively.
Since the $e^+e^-\to h_b(\text{1P},\text{2P})\pi^+\pi^-$ production dominantly proceeds via intermediate $Z_b(10610)$ and $Z_b(10650)$ states~\cite{Bondar:2011aa}, we require the recoil mass of one pion, $M^{\text{rec}}_{\pi^{\pm}}$, to be in the interval $(10.59, 10.67)\text{ GeV}/c^2$.

Next, we apply a set of requirements to select $h_{b}(\text{1P,2P}) \to \Upsilon(\text{1S})\eta(\pi^0)$ signal events.
To reconstruct the $\Upsilon({\text{1S}})\to \ell^+\ell^-$ decay, we require the presence of two oppositely charged leptons with an invariant mass within the $\pm 5\sigma$ interval $(9.235, 9.685)\text{ GeV}/c^2$, and with the same $dz$ and $dr$ requirements as for pions.
Muons are identified with a requirement on the likelihood ratio $\frac{\mathcal{L}_{\mu}}{\mathcal{L}_{\mu}+\mathcal{L}_{\pi}+\mathcal{L}_{K}}>0.1$, where the likelihood $\mathcal{L}_{i}$ ($i= \mu,\pi,K$) is assigned based on the length of the extrapolated track in the KLM and on the deviation of KLM hit positions from the extrapolated track~\cite{Abashian:2002bd}.
Electrons are identified with a requirement $\mathcal{P}_{e}>0.9$. 
The lepton identification efficiency is $93\%$ in the $\mu^+\mu^-$ mode and $94\%$ in the $e^+e^-$ mode.
To correct the possible energy loss due to bremsstrahlung radiation and improve the invariant-mass resolution in the $e^+e^-$ mode, the total four-momentum of photons within $15$ mrad from the momentum direction of the electron is added to its four-momentum.

In addition, we define the mass difference $\delta M_{X}$ to be $\delta M_{X}=M_{\ell\ell X}-M_{\ell\ell}$, and expect $\delta M_{\gamma\gamma}$ to peak at the value ($m_{h_b(\text{1P,2P})}-m_{\Upsilon(\text{1S})}$) with a resolution of $9$ and $12$~MeV$/c^2$ for the $h_b(\text{1P})$ and $h_b(\text{2P})$ mode, respectively.
Therefore, we set a $\pm 3\sigma$ requirement on $\delta M_{\gamma\gamma}$.

To suppress background from the $\Upsilon(\text{5S})\to \Upsilon(\text{2S})[\to \Upsilon(\text{1S})\pi\pi]\eta$ decay that peaks at $\delta M_{\pi\pi}=563$~MeV/$c^2$, we require $\delta M_{\pi\pi}>581$~MeV/$c^2$ rejecting $98.5\%$ of background events while retaining $91.5\%$ of the signal.
We also require $M_{\pi\pi\gamma\gamma}>810~\text{MeV}/c^2$ to suppress the $e^+e^-\to \ell^+\ell^- q \Bar{q}$ process where the dominant contribution is from events with an $\omega$ meson in the final state that peaks at $M_{\pi\pi\gamma\gamma}=782~\text{MeV}/c^2$.
This requirement rejects $97\%$ of $e^+e^-\to \ell^+\ell^-\omega$ events, has no effect on the $\eta$ signal, and retains $90\%$ of the $\pi^0$ signal events.
To reduce background from photon conversion, we apply a requirement $\cos{ \theta_{\pi^+\pi^-}}<0.95$, where $\theta_{\pi^+\pi^-}$ is the opening angle between the charged pions in the laboratory frame.

For selected candidates, we perform a kinematic fit constraining the four-momentum of all final-state particles to that of the initial $e^+e^-$ system and apply a requirement on the fit $\chi^2$ value to be less than $100~[55]$ for the $h_b(\text{1P})$ [$h_b(\text{2P})$] mode.
Where there are multiple candidates in an event, the one with the lowest $\chi^2$ is retained.
The requirement on $\chi^2$ is optimized based on a figure-of-merit $S/\sqrt{S + B}$, where $S$ is the number of signal events expected assuming a $2\%$ branching fraction, and $B$ is the number of background events estimated in the signal region based on that observed in the $M^{\text{rec}}_{\pi\pi}$ sideband.
The $M^{\text{rec}}_{\pi\pi}$ signal regions are defined as $(9.881, 9.917)\text{ GeV}/c^2$ for the $h_b(\text{1P})$ and $(10.242, 10.278)\text{ GeV}/c^2$ for the $h_b(\text{2P})$, corresponding to a $3\sigma$ window.
The $M_{\gamma\gamma}$ signal regions are defined as $(110, 155)\text{ MeV}/c^2$ for the $\pi^0$ and $(450, 600)\text{ MeV}/c^2$ for the $\eta$, corresponding to a $5\sigma$ window around their known values~\cite{Workman:2022ynf}.
The $M^{\text{rec}}_{\pi\pi}$ sidebands are defined as $(9.75, 9.99)\text{ GeV}/c^2$ and $(10.19, 10.32)\text{ GeV}/c^2$, excluding the above signal regions.
These ranges do not include contributions from the $\Upsilon(\text{2S})$, $\Upsilon(\text{1D})$, and $\Upsilon(\text{3S})$ resonances that peak nearby.

After applying all selection requirements, we expect the remaining background to dominantly come from the $\Upsilon(\text{5S})\to\chi_{bJ}(\text{1P})[\to \Upsilon(\text{1S})\gamma]\pi^+\pi^-\pi^0$ decay.
Other bottomonium transitions with the emission of an $\eta$($\pi^0$) are either excluded by the $M^{\text{rec}}_{\pi\pi}$ range, for example, $\Upsilon(\text{5S})\to\Upsilon(\text{1D})\eta$, or have a negligible branching fraction, for example, $\Upsilon(\text{5S})\to\Upsilon(\text{2S,3S})[\to \Upsilon(\text{1S})\eta(\pi^0\pi^0)]\pi^+\pi^-$.
Processes with $B$-mesons are expected to make no contribution to the background.
A possible contamination from the not-yet-observed $h_b(\text{2P})\to \chi_{bJ}(\text{1P})[\to \Upsilon(\text{1S})\gamma]\gamma$ process is included in our fit model as described below.

The $h_b(\text{1P},\text{2P})$ reconstruction efficiency, $\varepsilon_{h_b}$, is determined from MC simulation as $N_{\text{det}}/(N_{\text{gen}}\varepsilon_{\pi\pi})$, where $N_{\text{det}}$ is the MC signal yield, $N_{\text{gen}}$ is the number of generated events, and $\varepsilon_{\pi\pi}$ is the $\pi^+\pi^-$ reconstruction efficiency determined from an MC sample where only the pion pair is reconstructed following the requirements of Ref.~\cite{Belle:2012fkf}.
All selection criteria and $h_b(\text{1P},\text{2P})$ reconstruction efficiencies are summarized in Table~\ref{tab:selection_criteria}.

\begin{table}[!htb]
\footnotesize
\caption{Selection criteria and $h_b(\text{1P},\text{2P})$ reconstruction efficiencies.}
    \label{tab:selection_criteria}
\begin{tabular}
{l c c c }
\hline \hline
Criterion & $h_b(\text{2P})\to\Upsilon(\text{1S})\eta(\pi^0)$ &   $h_b(\text{1P})\to\Upsilon(\text{1S})\pi^0$ \\
        \hline
        $M^{\text{\text{rec}}}_{\pi^{\pm}}$ (GeV/$c^2$) & $(10.59, 10.67)$ & $(10.59, 10.67)$ \\
        $M_{\ell\ell}$ (GeV/$c^2$)   & $(9.235,~9.685)$ & $(9.235,~9.685)$ \\
        $\delta M_{\gamma\gamma}$ (MeV/$c^2$) &  $(761,~833)$  & $(412,~466)$ \\ 
        $\delta M_{\pi\pi}$ (MeV/$c^2$) & $>581$ & -- \\
        $M_{\pi\pi\gamma\gamma}$ (MeV/$c^2$) & $>810$ & $>810$ \\
        $\cos{\theta_{\pi\pi}}$  & $<0.95$ & $<0.95$ \\
        $\chi^2$  & $<55$  & $<100$  \\
        \hline
        $\varepsilon^{\mu\mu}_{h_b}$ ($\%$) & $33.1 \pm 0.3$ ($25.6 \pm 0.2$) &  $24.8 \pm 0.2$ \\
        $\varepsilon^{ee}_{h_b}$ ($\%$) & $23.7\pm 0.2$ ($18.8 \pm 0.2$) & $18.7\pm 0.2$ \\

\hline \hline
\end{tabular}
\end{table}

\begin{figure}[!t]
  \centering
  \includegraphics[width=1\linewidth]{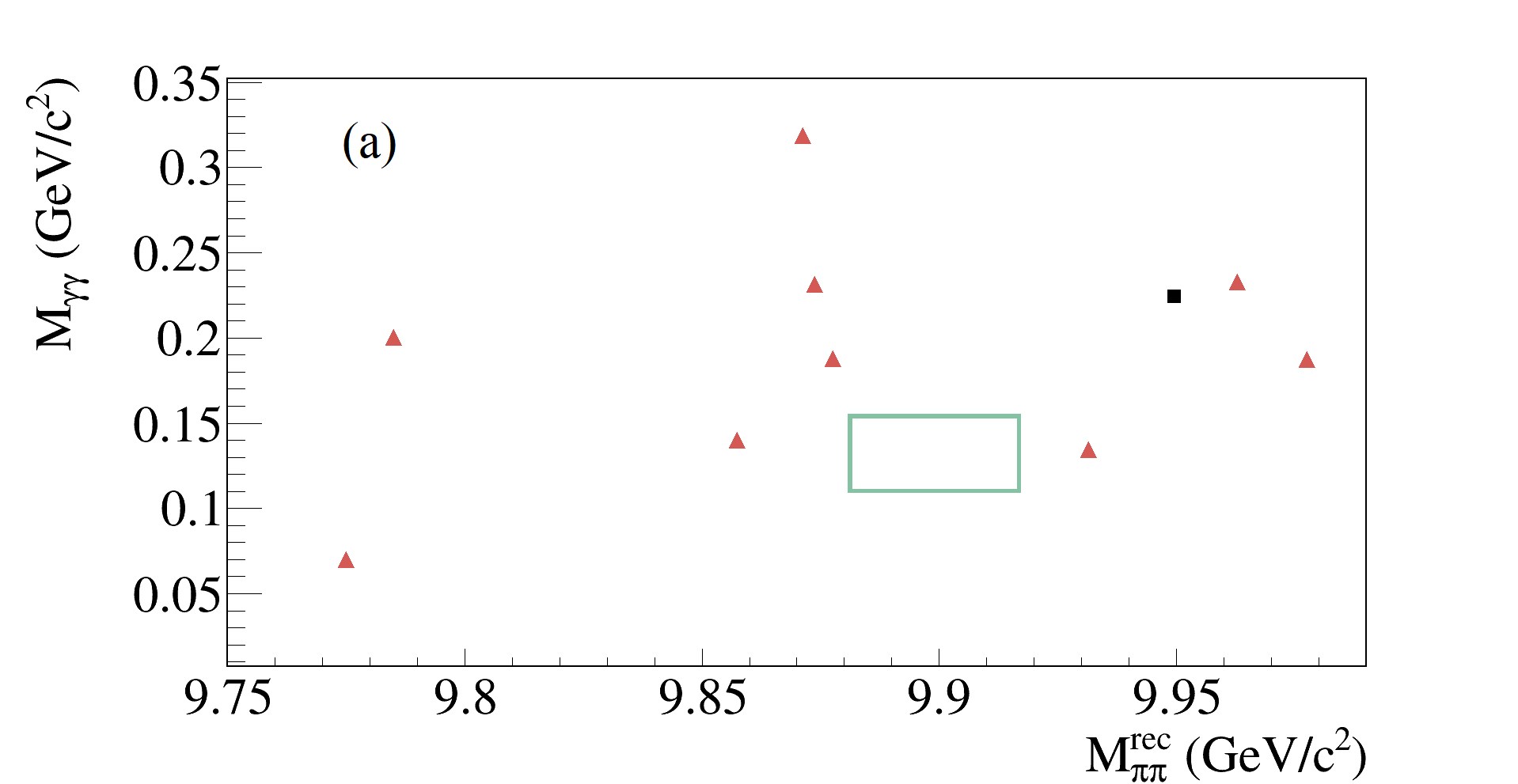} \hfill
  \includegraphics[width=1\linewidth]{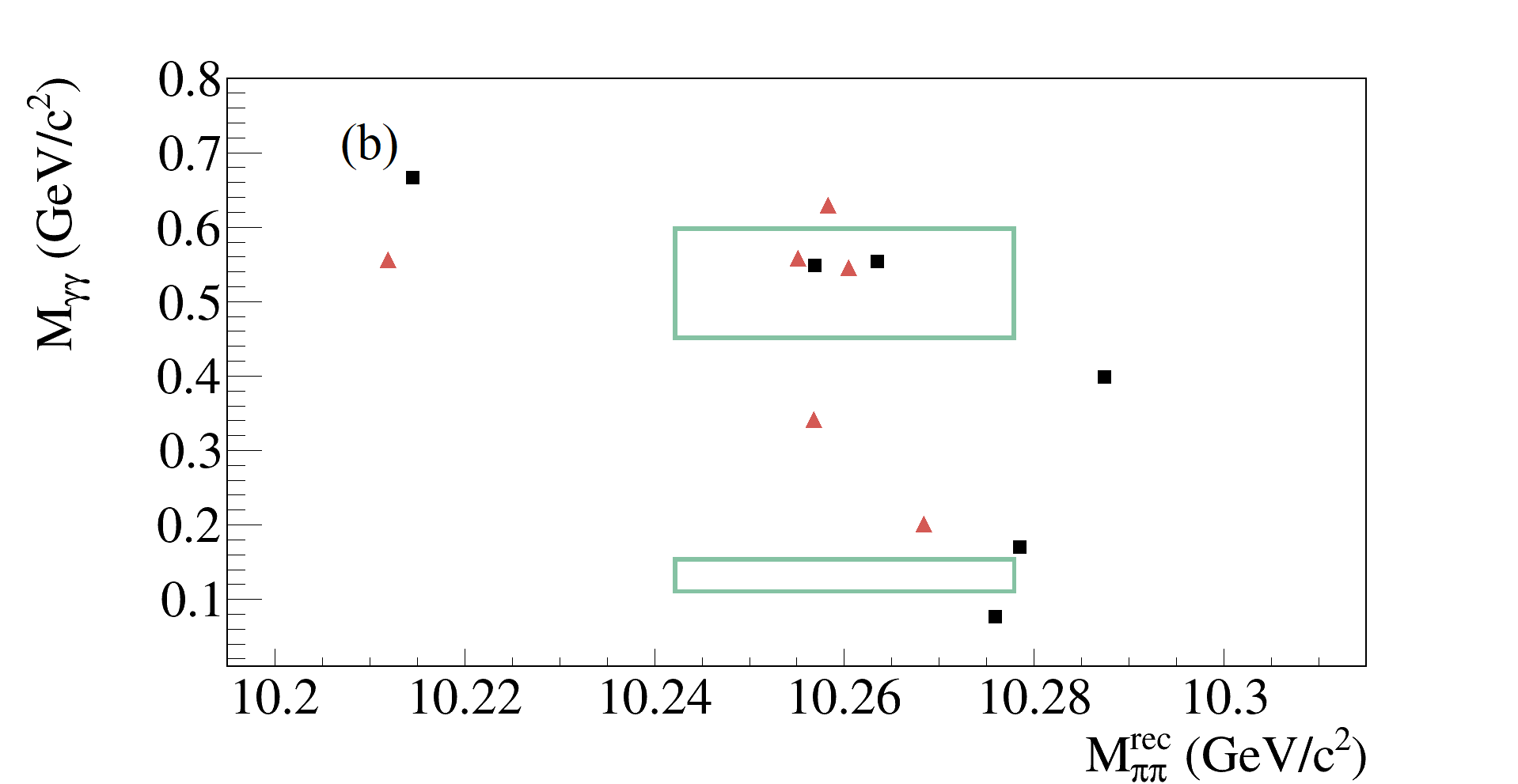} \\
  \caption{The $M_{\gamma\gamma}$ vs $M^{\text{rec}}_{\pi\pi}$ distribution for the $h_b(\text{1P})$ (a) and $h_b(\text{2P})$ (b) modes in the data. Black squares are for the $\mu^+\mu^-$ data and red triangles for the $e^+e^-$ data. Green rectangles represent the signal regions.}
    \label{img:exp_data}
\end{figure}

Figure~\ref{img:exp_data} displays the $M_{\gamma\gamma}$ vs. $M^{\text{rec}}_{\pi\pi}$ distribution for the selected events in the data.
We combine the $\mu^+\mu^-$ and $e^+e^-$ modes to determine the signal yield.
As we find no events in the signal regions of the $h_b(\text{1P,2P})\to\Upsilon(\text{1S})\pi^0$ processes, we set upper limits on the corresponding branching fractions using a counting method described below.

For the $h_b(\text{2P})\to\Upsilon(\text{1S})\eta$ decay, the signal yield is determined from an unbinned extended maximum-likelihood fit to the $M_{\gamma\gamma}$ vs. $M^{\text{rec}}_{\pi\pi}$ distribution.
The fit function is the sum of a signal component, a combinatorial background component, comprising both $\pi^0$ and non-$\pi^0$ events, and a component of the $\Upsilon(\text{5S})\to\chi_{bJ}(\text{1P})[\to \Upsilon(\text{1S})\gamma]\pi^+\pi^-\pi^0$ process.

\begin{figure}[!t]
  \centering
  \includegraphics[width=1\linewidth]{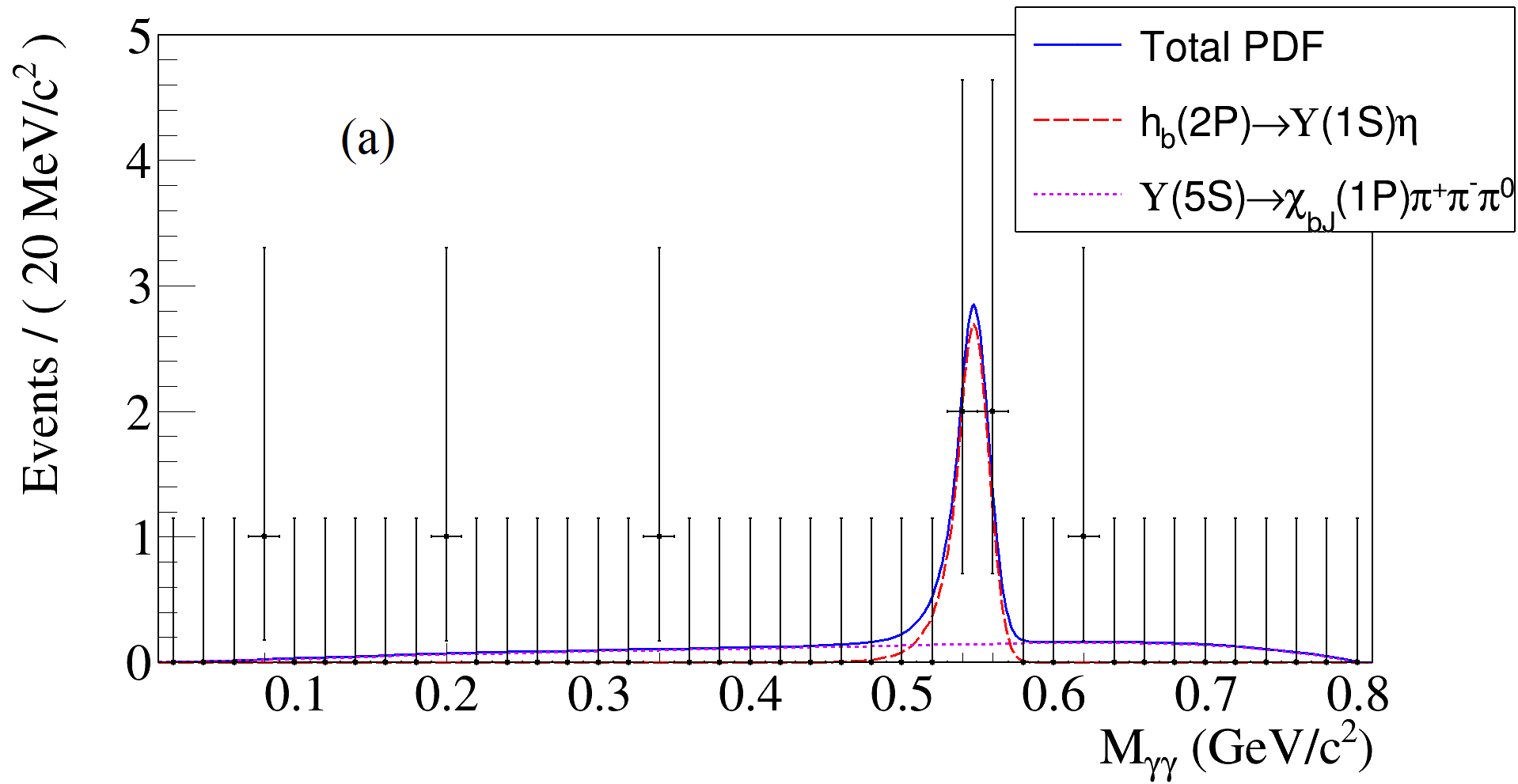} \hfill
  \includegraphics[width=1\linewidth]{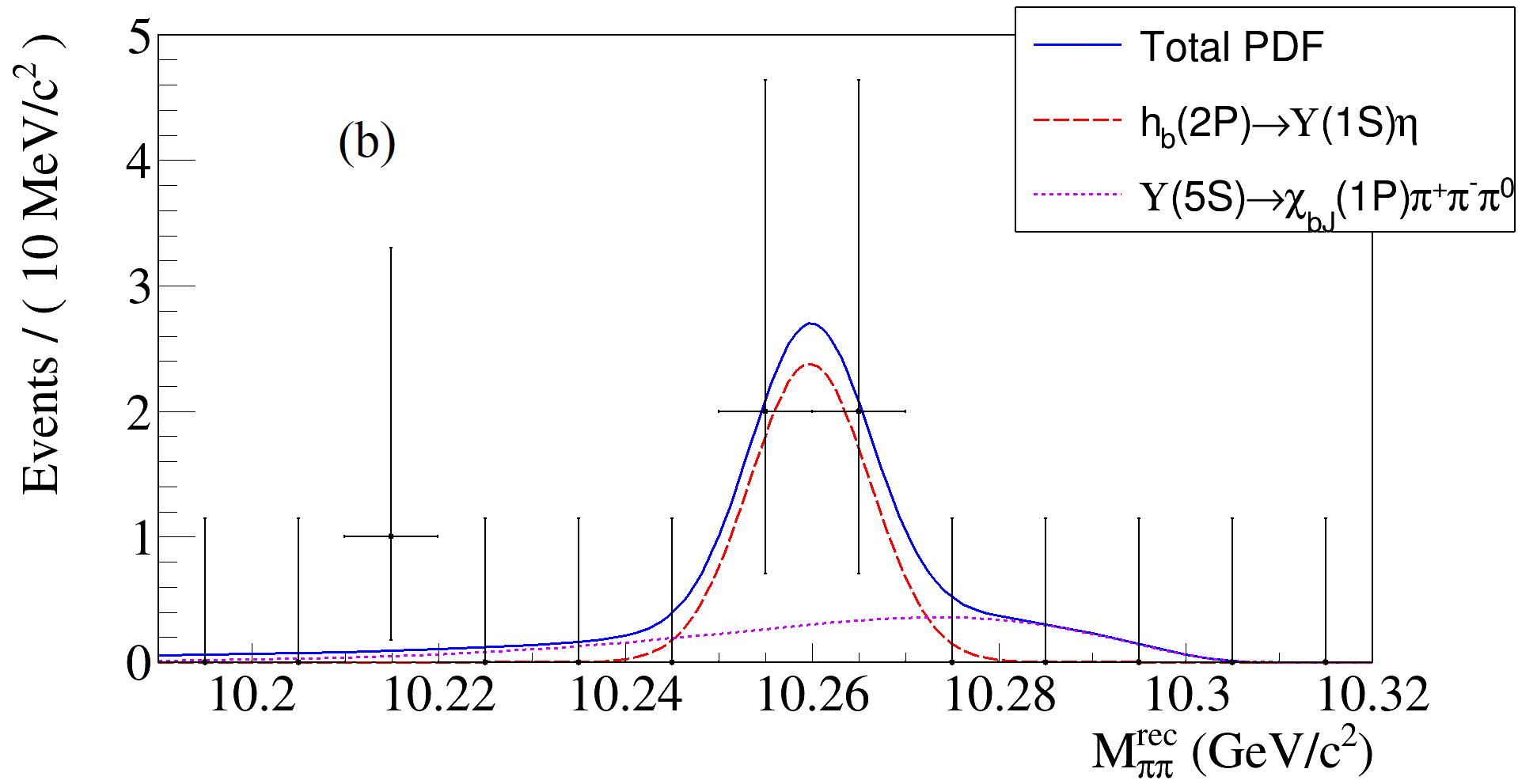} \\
  \caption{Comparison of results of the $M_{\gamma\gamma}$ vs $M^{\text{rec}}_{\pi\pi}$ fit (solid lines) with the data (points with error bars) shown as projection on (a) $M_{\gamma\gamma}$ with the additional requirement of $10.242 < M^{\text{rec}}_{\pi\pi} < 10.278~\text{GeV}/c^2$ and on (b) $M^{\text{rec}}_{\pi\pi}$ with the additional requirement of $450 < M_{\gamma\gamma} < 600~\text{MeV}/c^2$. Since both combinatorial background components are small, we do not plot them.}
    \label{img:fit_results}
\end{figure}

Shapes of all components are fixed from the dedicated MC samples, while their yields are floated.
The shape of the signal component is parameterized by a product of a Crystal Ball function~\cite{Skwarnicki:1986xj} in $M_{\gamma\gamma}$ and a Gaussian function in $M^{\text{rec}}_{\pi\pi}$.
For the background modeling, we use a threshold function $f(x)=(x-p_1)^{p_2}e^{p_3x}$.
Non-$\pi^0$ combinatorial background and $\Upsilon(\text{5S})\to\chi_{bJ}(\text{1P})[\to \Upsilon(\text{1S})\gamma]\pi^+\pi^-\pi^0$ components are separately parameterized as $f(M_{\gamma\gamma})\times f(M^{\text{rec}}_{\pi\pi})$.
The $\pi^0$ combinatorial background component is parameterized as a product of the Crystal Ball function in $M_{\gamma\gamma}$ and $f(M^{\text{rec}}_{\pi\pi})$.
Results of the fit are shown in Fig.~\ref{img:fit_results}.

We evaluate the significance as $\sqrt{2 \log{[\mathcal{L}(N_{\text{sig}})/\mathcal{L}(0)}]}$, where $\mathcal{L}(N_{\text{sig}})$ and $\mathcal{L}(0)$ are the likelihoods for a fit that includes a signal yield $N_{\text{sig}}$ and a fit with the background hypothesis only, respectively.   
The branching fraction, $\mathcal{B}$, is calculated as 
\begin{equation}
\label{eq:branching}
    \mathcal{B} =\frac{N_{\text{sig}}}{N^{\text{tot}}_{h_b(\text{1P},\text{2P})} (\mathcal{B}^{\mu\mu} \varepsilon^{\mu\mu}_{h_b} + \mathcal{B}^{ee} \varepsilon_{h_b}^{ee}) }, 
\end{equation}
where $N_{\text{sig}}$ is the signal yield, $N^{\text{tot}}_{h_b(\text{1P})} = (84.2 \pm 4.4 ^{+2.1} _{-1.3})\times10^{3}$, $N^{\text{tot}}_{h_b(\text{2P})} = (98.5 \pm 8.1 ^{+5.5} _{-6.3})\times10^{3}$ \cite{Belle:2012fkf}, $\mathcal{B}^{\ell\ell}$ is the product of the intermediate branching fractions for the process, and $\varepsilon^{\ell\ell}_{h_b}$ is the $h_b(\text{1P},\text{2P})$ reconstruction efficiency for the corresponding $\ell^+\ell^-$ mode.

The $h_b(\text{2P})\to\Upsilon(\text{1S})\eta$ signal yield is $3.8 ^{+2.0} _{-1.7}$, corresponding to $4.0\sigma$ significance and $\mathcal{B}[h_{b}(\text{2P}) \to \Upsilon(\text{1S})\eta]=(7.1 ~^{+3.7} _{-3.2}\pm0.8)\times10^{-3}$, where the first uncertainty is statistical and the second one is systematic uncertainty, which is described below.

We estimate systematic uncertainties of $0.7\%$ from the reconstruction efficiency of tracks, $2.0\%$ ($2.8\%$) from the identification of two muons (electrons), and $3.0\%$ from the reconstruction efficiency of photons.
The main uncertainty comes from the measurement of $N^{\text{tot}}_{h_b(\text{1P},\text{2P})}$ that is $5.6\%$ or $10.1\%$ respectively~\cite{Belle:2012fkf}.
For the $h_b(\text{2P})\to\Upsilon(\text{1S})\eta$ process, the estimated uncertainty of $3.0\%$ comes from varying the shapes of combinatorial background components, allowing parameters of $f(M^{\text{rec}}_{\pi\pi})$ to float or replacing it with a flat function.
To test the possible contribution from $h_b(\text{2P})\to \chi_{bJ}(\text{1P})[\to \Upsilon(\text{1S})\gamma]\gamma$, we perform the fit to data, adding the component parameterized as a product of a Gaussian in $M^{\text{rec}}_{\pi\pi}$ and $f(M_{\gamma\gamma})$ and fixed from the dedicated MC study.  
We add a Poisson term to the likelihood function with an expected number of events equal to that obtained from our fit to data and with zero observed events as found in the dedicated study~\cite{hbygam}, corrected for the efficiency ratio.
The fit yields a value consistent with zero for $h_b(\text{2P})\to \chi_{bJ}(\text{1P})\gamma$ and its effect is included as systematic uncertainty.
We find the signal significance for $h_b(\text{2P})\to\Upsilon(\text{1S})\eta$ to be $3.5\sigma$, taking the lowest value with the above systematic variations.

To set upper limits on the $h_b(\text{1P,2P})\to\Upsilon(\text{1S})\pi^0$ branching fractions, we use the Rolke approach~\cite{Rolke:2004mj} with bounded likelihood that takes into account an uncertainty in the background level and the multiplicative systematic uncertainty that is $6.5\%$ and $11.0\%$ for the $h_b(\text{1P})$ and $h_b(\text{2P})$, respectively. 
The background level is estimated from a fit to the data in which the $\Upsilon(\text{1S})\pi^0$ yield is set to zero; this finds $0.43\pm0.15$ and $0.10\pm0.05$ events for the $h_b(\text{1P})$ and $h_b(\text{2P})$, respectively.
We determine upper limits to be $\mathcal{B}[h_{b}(\text{1P})\to \Upsilon(\text{1S})\pi^0] < 1.8\times10^{-3}$ and $\mathcal{B}[h_{b}(\text{2P}) \to \Upsilon(\text{1S})\pi^0] < 1.8\times10^{-3}$ at $90\%$ confidence level.

The obtained $\mathcal{B}[h_{b}(\text{2P}) \to \Upsilon(\text{1S})\eta]=(7.1 ~^{+3.7} _{-3.2}\pm0.8)\times10^{-3}$ does not noticeably change the ratio $R_{h_b} / R_{\chi_{b1}}$ which only decreases from $0.24$ to $0.23$; however, it differs from the expectation that it could be as large as $10\%$.
The latter expectation comes from a comparison of the $h_{b}(\text{2P}) \to \Upsilon(\text{1S})\eta$ decay to a similar $\Upsilon(\text{3S})\to h_b(\text{1P})\pi^0$ decay -- the ratio of their widths is expected to be in range $140-320$~\cite{Li:2012is}.
To calculate $\Gamma[h_{b}(\text{2P}) \to \Upsilon(\text{1S})\eta]$ the total width $\Gamma_{h_{b}(\text{2P})}$ is required, which is not currently measured. 
However, there are many theoretical predictions for this quantity that predict it to be in the range of $50-100$~keV~\cite{Godfrey:2002rp, Ebert:2002pp, Segovia:2016xqb}. 
Using $\Gamma_{h_{b}(\text{2P})} = 75\pm25$~keV,  $\Gamma_{\Upsilon(\text{3S})}=20.32\pm1.85$~keV~\cite{Workman:2022ynf}, $\mathcal{B}[\Upsilon(\text{3S})\to h_b(\text{1P})\pi^0] \times \mathcal{B}[h_b(\text{1P}) \to \eta_b(\text{1S})\gamma)] = (4.3\pm1.1\pm0.9)\times 10^{-4}$~\cite{BaBar:2011ljf}, $\mathcal{B}[h_b(\text{1P}) \to \eta_b(\text{1S})\gamma)] = 49.2\pm5.7^{+5.6}_{-3.3}\%$~\cite{Belle:2012fkf}, and the obtained $\mathcal{B}[h_{b}(\text{2P}) \to \Upsilon(\text{1S})\eta]$, we calculate the ratio 
\begin{equation}
\label{eq:width_frac}
    \frac{ \Gamma[h_{b}(\text{2P}) \to \Upsilon(\text{1S})\eta] }{ \Gamma[\Upsilon(\text{3S})\to h_b(\text{1P}) \pi^0] } = 30^{+20}_{-18} \pm10, 
\end{equation}
where the first uncertainty is the combined uncertainty of the measurements and the second comes from $\Gamma_{h_{b}(\text{2P})}$. 
This result is noticeably lower than the expected range of $140$ to $320$. 
Such a discrepancy could be accounted for either by experimental uncertainties or by some unaccounted effects in the prediction model, such as an admixture of $B\bar{B}$ states in the bottomonium system. 

In summary, we have reported the first results on experimental studies of $h_{b}(\text{1P},\text{2P})$ decays into the $\Upsilon(\text{1S})$ with the emission of $\pi^0$ or $\eta$ mesons.
We find evidence for the $h_{b}(\text{2P}) \to \Upsilon(\text{1S})\eta$ transition, with $\mathcal{B}[h_{b}(\text{2P}) \to \Upsilon(\text{1S})\eta]=(7.1 ~^{+3.7} _{-3.2}\pm0.8)\times10^{-3}$. 
The signal significance is $3.5\sigma$, including systematic uncertainty.
In the absence of any signal for the isospin-violating decays $h_{b}(\text{1P,2P})\to \Upsilon(\text{1S})\pi^0$, we set upper limits on their branching fractions: $\mathcal{B}[h_{b}(\text{1P})\to \Upsilon(\text{1S})\pi^0] < 1.8\times10^{-3}$ and $\mathcal{B}[h_{b}(\text{2P}) \to \Upsilon(\text{1S})\pi^0] < 1.8\times10^{-3}$ at the $90\%$ confidence level.
The measured $\mathcal{B}[h_{b}(\text{2P}) \to \Upsilon(\text{1S})\eta]$ value is noticeably smaller than the prediction of $10\%$~\cite{Li:2012is}.
Further efforts are necessary in order to clarify the situation.
An additional round of data taking at the $\Upsilon(\text{5S})$ energy with Belle~II~\cite{Belle-II:2018jsg} would allow hadronic transitions within the bottomonium family to be studied with a new level of precision.


\section{Acknowledgement}

This work, based on data collected using the Belle detector, which was
operated until June 2010, was supported by 
the Ministry of Education, Culture, Sports, Science, and
Technology (MEXT) of Japan, the Japan Society for the 
Promotion of Science (JSPS), and the Tau-Lepton Physics 
Research Center of Nagoya University; 
the Australian Research Council including grants
DP210101900, 
DP210102831, 
DE220100462, 
LE210100098, 
LE230100085; 
Austrian Federal Ministry of Education, Science and Research (FWF) and
FWF Austrian Science Fund No.~P~31361-N36;
National Key R\&D Program of China under Contract No.~2022YFA1601903,
National Natural Science Foundation of China and research grants
No.~11575017,
No.~11761141009, 
No.~11705209, 
No.~11975076, 
No.~12135005, 
No.~12150004, 
No.~12161141008, 
and
No.~12175041, 
and Shandong Provincial Natural Science Foundation Project ZR2022JQ02;
the Czech Science Foundation Grant No. 22-18469S;
Horizon 2020 ERC Advanced Grant No.~884719 and ERC Starting Grant No.~947006 ``InterLeptons'' (European Union);
the Carl Zeiss Foundation, the Deutsche Forschungsgemeinschaft, the
Excellence Cluster Universe, and the VolkswagenStiftung;
the Department of Atomic Energy (Project Identification No. RTI 4002), the Department of Science and Technology of India,
and the UPES (India) SEED finding programs Nos. UPES/R\&D-SEED-INFRA/17052023/01 and UPES/R\&D-SOE/20062022/06; 
the Istituto Nazionale di Fisica Nucleare of Italy; 
National Research Foundation (NRF) of Korea Grant
Nos.~2016R1\-D1A1B\-02012900, 2018R1\-A2B\-3003643,
2018R1\-A6A1A\-06024970, RS\-2022\-00197659,
2019R1\-I1A3A\-01058933, 2021R1\-A6A1A\-03043957,
2021R1\-F1A\-1060423, 2021R1\-F1A\-1064008, 2022R1\-A2C\-1003993;
Radiation Science Research Institute, Foreign Large-size Research Facility Application Supporting project, the Global Science Experimental Data Hub Center of the Korea Institute of Science and Technology Information and KREONET/GLORIAD;
the Polish Ministry of Science and Higher Education and 
the National Science Center;
the Ministry of Science and Higher Education of the Russian Federation
and the HSE University Basic Research Program, Moscow; 
University of Tabuk research grants
S-1440-0321, S-0256-1438, and S-0280-1439 (Saudi Arabia);
the Slovenian Research Agency Grant Nos. J1-9124 and P1-0135;
Ikerbasque, Basque Foundation for Science, and the State Agency for Research
of the Spanish Ministry of Science and Innovation through Grant No. PID2022-136510NB-C33 (Spain);
the Swiss National Science Foundation; 
the Ministry of Education and the National Science and Technology Council of Taiwan;
and the United States Department of Energy and the National Science Foundation.
These acknowledgements are not to be interpreted as an endorsement of any
statement made by any of our institutes, funding agencies, governments, or
their representatives.
We thank the KEKB group for the excellent operation of the
accelerator; the KEK cryogenics group for the efficient
operation of the solenoid; and the KEK computer group and the Pacific Northwest National
Laboratory (PNNL) Environmental Molecular Sciences Laboratory (EMSL)
computing group for strong computing support; and the National
Institute of Informatics, and Science Information NETwork 6 (SINET6) for
valuable network support.

\bibliography{ekoval}

\end{document}